# Lookahead and Hybrid Sample Allocation Procedures for Multiple Attribute Selection Decisions

Jeffrey W. Herrmann, Kunal Mehta

Department of Mechanical Engineering and Institute for Systems Research

University of Maryland

**Abstract**

Attributes provide critical information about the alternatives that a decision-maker is considering. When their magnitudes are uncertain, the decision-maker may be unsure about which alternative is truly the best, so measuring the attributes may help the decision-maker make a better decision. This paper considers settings in which each measurement yields one sample of one attribute for one alternative. When given a fixed number of samples to collect, the decision-maker must determine which samples to obtain, make the measurements, update prior beliefs about the attribute magnitudes, and then select an alternative. This paper presents the sample allocation problem for multiple attribute selection decisions and proposes two sequential, lookahead procedures for the case in which discrete distributions are used to model the uncertain attribute magnitudes. The two procedures are similar but reflect different quality measures (and loss functions), which motivate different decision rules: (1) select the alternative with the greatest expected utility and (2) select the alternative that is most likely to be the truly best alternative. We conducted a simulation study to evaluate the performance of the sequential procedures and hybrid procedures that first allocate some samples using a uniform allocation procedure and then use the sequential, lookahead procedure. The results indicate that the hybrid procedures are effective; allocating many (but not all) of the initial samples with the uniform allocation procedure not only reduces overall computational effort but also selects alternatives that have lower average opportunity cost and are more often truly best.

This work is published on arXiv under the Creative Commons Attribution-NonCommercial-NoDerivatives 4.0 International license (https://creativecommons.org/licenses/by-nc-nd/4.0/).

**1. Introduction**

In a multiple attribute selection decision, a finite set of alternatives is given, and the alternatives are described by several attributes that are important to the decision-maker, who will pick an alternative. (Attributes are also known as *evaluation measures* or *performance scores*.) To compare the alternatives and identify the best one, the decision-maker may model his preferences to develop a value function and select the alternative that has the greatest value (Keeney, 1974; Keeney and Raiffa, 1993). If the magnitudes of the attributes are known (that is, there is no uncertainty about them), then this is straightforward. If the magnitudes of the attributes are unknown, however, then the decision-maker has a more difficult problem. (Here, the term "magnitude" is used to describe the length, mass, score, level, or other quantity that is associated with the attribute.)

For example, consider the following two selection decisions. In the first, the Domestic Nuclear Detection Office (DNDO) was evaluating radiation detection systems to be installed in U.S. airports (Leber, 2016). This office was considering 576 different system designs. The value of any system design was a function of multiple attributes, including the system's ability to detect eleven different types of radiological or nuclear material. Given a limited set of resources for testing these systems, the office had to determine how many times to test the different systems on each of the eleven types of material to obtain estimates of the magnitudes of these attributes and select the best system.

Consider also the selection of an aluminum alloy to replace the gray cast iron used in a part for a motor. The alternatives include multiple alloys that can be processed in different ways. The following properties are important for making a decision among alloy alternatives: yield strength, ultimate tensile strength, shear strength, and elongation (Dieter and Schmidt, 2013). Although typical magnitudes are available from materials handbooks, the magnitudes of these properties for these alloys when made into the desired part is unknown. Given the opportunity to obtain measurements of these properties from test parts that have been manufactured, the decision-maker needs to determine which set of measurements will most help the decision-making select the best alloy. Each measurement returns the magnitude of one property (attribute) of one alloy (alternative). Thus, each measurement is incomplete information about that alternative (Lawrence, 1999). The measurements are not perfectly accurate, however; the magnitude returned may contain some error. After obtaining these measurements, the decision-maker can update his beliefs about these properties for these alloys and select the alloy that appears to be the best.

This paper considers a decision-maker who will acquire a fixed amount of information about the attributes' magnitudes, which will reduce the attribute uncertainty, in order to improve the expected quality of the alternative that is selected after the information has been obtained (the amount of information is fixed due to financial, time, or resource constraints). We assume that there are true magnitudes for every attribute of every alternative, but these are unknown. The information that the decision-maker will gather is imperfect because it is the output of measurement processes that have random error. Each sample is the result of an experiment that measures one attribute of one alternative. The sample is the true magnitude of that attribute plus the measurement error. After determining which samples to obtain and reviewing the sample results, the decision-maker will select the alternative that is most preferred.

Determining the expected monetary value of the information that a sample produces is one way to determine which samples to obtain, and that is appropriate if the decision-maker measures value on a monetary scale. Because money is not the only way to measure value, however, using a monetary value is not necessary for making this decision, and the problem formulation (Section 3.2) uses a more general approach that allows but does not require a monetary value.

Determining which samples to obtain is a metareasoning decision. This metareasoning problem is a version of the decision problem analyzed by Raiffa and Schlaifer (1961). In this problem, each experiment can measure only one attribute of one alternative, and the decision-maker has a fixed number of experiments (measurements). The decision-maker wants to determine the best set of experiments (the best allocation of a fixed number of samples). Although previous work on ranking and selection has studied the problem of optimally allocating a computing budget (as discussed in Section 2), that work has focused on sampling different alternatives, where a sample returns the scores of one or more objectives for the sampled alternative. The problem studied here involves an additional aspect, for each sample returns a magnitude for only one attribute for one alternative.

This work builds on the work of Leber and Herrmann (2016), but the study described here considered attributes with discrete magnitudes, more general distributions, two different quality measures (loss functions), two versions of the sequential look-ahead procedure, and a novel hybrid sample allocation procedure that has not been previously proposed or evaluated. Although a sequential procedure can determine the best single sample to obtain next, this paper considers the question of whether a hybrid procedure (which first uses a simple sample allocation rule and then employs sequential sampling) can be more effective when many samples will be taken.

Motivated by scenarios in which the attributes are physical quantities that can be measured, this study considered the case in which all attribute magnitudes and samples are elements of finite discrete sets. From a practical matter, any measurement is an approximation of the true magnitude, and a measurement process has a finite resolution so that the measured magnitude is on a discrete scale. For example, if one measures the mass of an object on a 500-gram scale with a 0.1-gram resolution, the scale must quantize the measured mass, which yields a quantity on the discrete scale from 0.0g to 499.9g (Tutelman, 2007).

The problem with discrete magnitudes is very general, so no assumptions about the error distributions or the distributions that model the decision-maker's beliefs are needed. Using discrete magnitudes simplifies some of the mathematics involved, but it can require numerous computations to determine and update distributions. Because of this, we also considered a hybrid procedure that reduces the overall computational effort by including a batch allocation step (in which the samples are allocated uniformly among the alternatives and attributes).

This paper presents the sample allocation approaches and describes the results of a study of their performance on instances of the sample allocation problem. The study goals were to compare the sequential sample allocation procedures and hybrid procedures to determine how well they selected high-quality alternatives.

Section 2 provides a brief review of the related literature. Section 3 formulates the sample allocation problem. Section 4 presents the sample allocation procedures. Section 5 describes the experiment that was conducted. Section 6 discusses the results of the experiment. Section 7 concludes the paper.

**2. Literature Review**

Generally, this work builds on the foundations of multi-attribute decision analysis (Keeney and Raiffa, 1976, 1993), statistical decision theory (Raiffa and Schlaifer, 1961; Pratt *et al.*, 1995), optimization under uncertainty (Powell, 2016), and the economic value of information (Lawrence, 1999). Howard (1970) illustrated the key concepts related to analyzing experimentation decisions with a coin-flipping example. The approach proposed herein is related to the ranking and selection methods that use the outputs of discrete-event simulation or another stochastic process to estimate the performance of a finite number of alternatives (Kim and Nelson, 2006; Chen and Lee, 2011; Powell and Ryzhov, 2012; Powell, 2016).

Optimal computing budget allocation (OCBA) is a sequential approach for determining which alternatives should be sampled (e.g., Lee *et al*., 2010a, b; Teng *et al*., 2010; Chen and Lee, 2011). Each step samples one or more alternatives and uses the results to determine the next alternative(s) to sample; this is repeated until the number of samples reaches a specified limit (the computing budget). The indifference zone (IZ) approach (Kim and Nelson, 2006) seeks to minimize the number of samples needed to guarantee that the likelihood of selecting the truly best alternative is sufficiently great.

The expected value of the sample information (Raiffa and Schlaifer, 1961; Lawrence, 1999) measures the expected gain in the decision-maker's utility from the next sample (experiment). When the decision-maker can relate utility to monetary values, the value of the information has a clear economic meaning. In other cases, when the gain is measured abstractly (e.g., as an increase in utility, also known as the *utility increment* (Lawrence, 1999; cf. Bernardo, 1979)), the decision-maker can use it to compare different sources of information and select the best one. (Other measures have been proposed; for instance, Lindley, 1956, defined the "amount" of information provided by an experiment as the expected reduction in the entropy of the uncertain quantity.) Chick and Inoue (2001) presented two-stage and sequential approaches that seek to minimize the expected loss after the samples are obtained. Chick (2006) used the term "Value of Information Procedure (VIP)" to describe such approaches and emphasized these improve the expected value of the sample information. Procedures were presented for both opportunity cost and 0-1 loss. Branke (2007) compared OCBA, IZ, and VIP approaches and concluded that a version of OCBA was best for maximizing the probability of selecting a "good" alternative (one that is close to optimal) and that versions of OCBA and VIP approaches were best for minimizing expected opportunity cost.

In this line of research, each sample measures one or more objective functions for one alternative; the allocation does not have to determine which attribute (or objective function) to sample. This is the key difference with the problem considered herein, and the existing OCBA, IZ, and VIP approaches cannot be directly applied to this problem. Moreover, existing approaches assume that the beliefs about the performance measure are normally distributed, but the proposed approach does not.

Leber and Herrmann (2013, 2014a, b, 2015, 2016) studied some specific cases of the sample allocation problem considered in this study. Those studies generally assumed that the error distributions are normally distributed and that the decision-maker's beliefs about the unknown attribute magnitudes can be expressed as normal distributions. Leber and Herrmann (2016) and Leber (2016) presented a sequential sample allocation approach for that case and showed that, compared with simple sample allocation procedures, it increased the likelihood of selecting a truly best alternative (that is, it reduced the expected 0-1 loss). The study described herein modified that approach for the case of general discrete distributions and considered a second measure of decision quality.

Although much of the previous work has considered the case in which the samples from different alternatives are independently distributed, some researchers have considered the case in which correlation exists; see, for example, Fu (2007), Frazier *et al*. (2009, 2011), and Qu (2011, 2012).

Powell (2016) reviewed numerous techniques for optimization under uncertainty. Although it is an approximation, a lookahead model is a way to find a policy for a stochastic optimization problem. Lookahead models may use a limited time horizon, discretization, and other types of approximations. The sequential approach considered in this study is a type of lookahead model.

## 3. Selection and Sample Allocation

The decision-maker must allocate the samples, collect the sample results, update his beliefs about the unknown attribute magnitudes and the unknown utility of the alternatives, and use a selection rule (decision rule) to choose an alternative. This section describes the problem formulation and selection rules that were considered in this study.

### 3.1. Preliminaries

In this problem, the decision-maker does not need to determine whether collecting the samples is a worthwhile use of resources (such as money, time, or testing equipment). The key decision is which alternatives and attributes to sample with the resources that have been allocated. Every sample requires the same use of resources (has the same cost), but this cost is not relevant to the sample allocation problem.

The decision-maker's objective is to improve the expected quality of the alternative that is selected after the samples have been obtained. We considered two different measures of "quality": (I) the expected utility of the selected alternative and (II) the likelihood that the selected alternative is one of the truly best alternatives. These two measures correspond to two loss functions that measure the distance between a selected alternative and the optimal: (I) the opportunity cost and (II) the 0-1 loss (cf. Pratt *et al.*, 1995; Chick and Inoue, 2001; Frazier *et al.*, 2008). Maximizing expected quality is equivalent to minimizing expected loss. Section 3.2 explains the quality measures in detail.

Naturally, the sample allocation problem resembles the problem of maximizing the expected value of information, which is the expected gain in the decision-maker's well-being that results from getting the samples (Lawrence, 1999). The problem formulation in Section 3.2 can be used when the decision-maker has a value function that maps a vector of attribute magnitudes into a monetary value. Moreover, if the decision-maker is risk-neutral, then maximizing the expected value of the selected alternative (minimizing the expected opportunity cost) is equivalent to maximizing the expected monetary value of the samples. (If the decision-maker is risk-averse, then one can determine the expected monetary value of the samples by taking the difference in the certainty equivalents.)

The problem formulation is general enough, however, to include a range of situations, including those where the decision-maker's value function does not correspond to a monetary value. For the more general setting, the concept of *utility increment*, the expected gain in utility, is a feasible, but non-traditional, way to measure the expected value of the information (Lawrence, 1999).

### 3.2. Problem Formulation

This section introduces the notation and formulates the sample allocation problem.

Let $m$ be the number of distinct alternatives. Let $\{a_1,\ldots,a_m\}$ be the set of alternatives. (Although this numbering is arbitrary, it creates an ordering that is used to break ties when needed.) Let $k \geq 2$ be the number of attributes that the decision-maker is considering. Let $\mu_{ij}$ be the actual magnitude (e.g., length, mass, score, or other quantity) of attribute $j$ of alternative $a_i$; this magnitude is unknown to the decision-maker. For alternative $a_i$, the vector of attribute magnitudes $(\mu_{i1}, \ldots, \mu_{ik})$ is a point in the consequence space (Keeney and Raiffa, 1976). Thus, the uncertain "state of the world" is the vector of attribute magnitudes for every alternative: $(\mu_{11}, \ldots, \mu_{mk})$.

The value functions reflect the decision-maker's preferences about the attributes and together map a vector of attribute magnitudes to a value as follows. For $j = 1, \ldots, k$, let $X_j$ be the set of possible magnitudes for attribute $j$ (across all of the alternatives), and let $v_j(x_j)$ be the single-attribute value function for attribute $j$, defined for all magnitudes $x_j \in X_j$. Let $\mathbf{X} = X_1 \times \cdots \times X_k$ be the consequence space. For all vectors $x = (x_1, \ldots, x_k) \in \mathbf{X}$, let $v(x) = f(v_1(x_1), \ldots, v_k(x_k))$ be a value function that combines the single-attribute value functions. This value could be located on a monetary scale (as discussed in Section 3.1), but that is not required. We make no assumptions about the form of the value functions. The computational experiment described in Section 5 considered both linear and nonlinear value functions.

Let $U(v)$ be the decision-maker's utility function, which expresses the degree of risk aversion with respect to uncertain value. Let $\xi_i = U(v(\mu_{i1}, \ldots, \mu_{ik}))$ be the true utility of alternative $a_i$. Let $\xi^* = \max_{i=1, \ldots, m} \{\xi_i\}$ and $A^* = \{a_i : \xi_i = \xi^*\}$. We make no assumptions about the form of the utility function. The computational experiment described in Section 5 considered utility functions for both risk-neutral and risk-averse decision-makers.

Let $T$ be the total number of samples that will be obtained (this is given). Let $N_{ij}$ be the total number of samples for attribute $j$ of alternative $a_i$ (all $N_{ij}$ are non-negative integers). The decision-maker's problem is to determine the $N_{ij}$ such that $\sum_{i=1}^{m}\sum_{j=1}^{k} N_{ij} = T$. Let $\varepsilon_j$ be the random error in the measurement process for attribute $j$. Let $E_j$ be the set of possible error quantities for the measurement process for attribute $j$. Let $p_j^e(e)$ be the probability that the random error in the measurement process for attribute $j$ equals $e$, for all $e$ in $E_j$. Let $W_{ij} = \mu_{ij} + \varepsilon_j$ be a random sample obtained from measuring attribute $j$ of alternative $a_i$; note that $W_{ij} \in X_j$. Information about measurement errors may be obtained from various sources, such as previous measurement data (including calibration data), knowledge about the measurement instrument, the specifications published by the instrument's manufacturer, and the uncertainties in published reference data (NIST, 2016).

Let $S_0$ be initial state of the decision-maker's beliefs about the attribute magnitudes; that is, $S_0$ is a set of prior distributions, one for each attribute for each alternative. Let $p_{ij}^0(x_j)$ be the prior distribution for attribute $j$ of alternative $a_i$. Let $S_t$ be state of the decision-maker's beliefs about the attribute magnitudes after obtaining $t$ samples $\{w_1,\ldots,w_t\}$. (A sample is a measurement of the magnitude of one attribute for one alternative.) This is a set of posterior distributions, one for each attribute for each alternative. (Section 3.3 presents the details of the state update.) We assume that the decision-maker's beliefs about the attributes and alternatives are independent. (The approach could be generalized to cover cases in which these beliefs are not independent, but that case was not considered in this work.) Because the decision-maker does not know the true attribute magnitudes, let $Y_{ij}^t$ be the random variable that represents the decision-maker's beliefs, after obtaining $t$ samples, about the magnitude of attribute $j$ of alternative $a_i$. Then, $p_{ij}^t(x_j) = P\{Y_{ij}^t = x_j\}$ for all $x_j \in X_j$. (These subjective probabilities are the decision-maker's beliefs about these magnitudes.)

Let $\Omega$ be the set of possible utilities (this is a discrete set because the attribute magnitudes are limited to discrete sets). Let $Z_i^t$ be the random variable that represents the decision-maker's beliefs, after obtaining $t$ samples, about the uncertain utility of alternative $a_i$. For all $z \in \Omega$, let

$$\tilde{p}_i^t(z) = P\{Z_i^t = z\}.$$

Define "$\succ$" as a relationship between $Z_i^t$ and $Z_h^t$ such that $Z_i^t \succ Z_h^t$ if and only if $Z_i^t > Z_h^t$ or ( $Z_i^t = Z_h^t$ and $i < h$). This relationship defines a rule for breaking ties. If $Z_i^t \succ Z_h^t$, then we say that alternative $i$ has a greater utility than alternative $h$.

Given the beliefs in state $S_t$, let $P_i^t$ be the probability that alternative $i$ has the greatest utility.

$$P_i^t = P\{Z_i^t \succ Z_h^t \ \forall h \neq i \,|\, S_t\} \tag{1}$$

As discussed in Section 3.1, we considered two measures of decision "quality": (I) the expected utility of the selected alternative and (II) the likelihood that the selected alternative is one of the truly best alternatives.

With quality measure I, the decision-maker's objective is equivalent to maximizing the expected value of the information gained from all of $T$ samples, which is equivalent to a monetary gain if the decision-maker's value function yields a monetary value; more generally, the gain can be measured as a utility increment that provides a non-monetary measure of the value of information.

The sample allocation problem can be formulated as follows:

$$\max E[\xi_b] \quad (2a)$$

$$\text{subject to } b = \underset{i=1,\ldots,m}{\arg\max} E\left[Z_i^T \mid w_1, \ldots, w_T\right] \quad (2b)$$

$$\sum_{i=1}^{m} \sum_{j=1}^{k} N_{ij} = T \quad (2c)$$

$$N_{ij} \in \{0,1,\ldots,T\} \text{ for } i=1,\ldots,m, \ j=1,\ldots,k \quad (2d)$$

With quality measure II, the decision-maker seeks to maximize the likelihood that the selected alternative $a_b$ is a truly best alternative (that is, $a_b \in A^*$). The decision-maker will select, after collecting $T$ samples, an alternative with the greatest probability $P_i^T$. In this case, the sample allocation problem can be formulated as follows:

$$\max P\{a_b \in A^*\} \quad (3a)$$

$$\text{subject to } b = \arg\max\{P_i^T : i=1,\ldots,m\} \quad (3b)$$

$$P_i^T = P\{Z_i^T \succ Z_h^T \ \forall h \neq i \mid S_T\} \text{ for } i=1,\ldots,m \quad (3c)$$

$$\sum_{i=1}^{m}\sum_{j=1}^{k} N_{ij} = T \qquad (3d)$$

$$N_{ij} \in \{0,1,\ldots,T\} \text{ for } i = 1,\ldots,m, \ j = 1,\ldots,k \qquad (3e)$$

Both formulations of the problem are stochastic because the obtained samples are random and thus $a_b$, the selected alternative, is random.

It is possible to consider the problem as a dynamic programming problem (cf. Frazier *et al.*, 2008), but the size of the state space (which is approximately $\left(mk \max_{j=1,\ldots,k} |X_j|\right)^t$ after obtaining $t$ samples) makes finding an optimal solution computationally intractable for even small problems. Thus, we studied sequential (lookahead) allocation procedures and hybrid allocation procedures that begin with a batch allocation and then continue with a sequential allocation procedure.

### 3.3. State Updates

The state is updated in a sequential manner after each sample is obtained. That is, after obtaining the *t*-th sample (from attribute *j* of alternative $a_i$), the decision-maker updates his beliefs. In the new state $S_t$, the probability distributions for almost all of the attribute magnitudes are the same as they were in the old state $S_{t-1}$. Only the probability distribution for the attribute that was sampled is updated using Bayes' rule.

Recall that $p_{ij}^t(x_j)$ is the probability, after obtaining $t$ samples, that the magnitude of attribute $j$ of alternative $a_i$ equals $x_j$. Let $w$ be the magnitude of the (just-obtained) sample of attribute $j$ of alternative $a_i$.

$$
\begin{aligned}
p_{ij}^t(x_j) &= P\{\text{attribute} = x_j \mid \text{sample} = w\} \\
&= \frac{P\{\text{attribute} = x_j \cap \text{sample} = w\}}{P\{\text{sample} = w\}} \\
&= \frac{p_{ij}^{t-1}(x_j) p_j^e(w - x_j)}{\sum_{y \in X_j} p_{ij}^{t-1}(y) p_j^e(w - y)}
\end{aligned}
\tag{4}
$$

The updated state also includes $\tilde{p}_i^t(z)$, the probability distributions for the alternative's utility. Let $x = (x_1, \ldots, x_k) \in \mathbf{X}$ be a vector that contains magnitudes for each attribute. Let $\bar{p}_i^t(x)$ be the probability (given state $S_t$) that this vector of magnitudes occurs for alternative $i$.

$$
\bar{p}_i^t(x) = \prod_{j=1}^{k} p_{ij}^t(x_j)
\tag{5}
$$

For $z \in \Omega$, let $\Psi(z)$ be the set of vectors $x \in \mathbf{X}$ with attribute magnitudes that yield (through the value function and the utility function) a utility of $z$:

$$
\Psi(z) = \{x : U(v(x)) = z\}
\tag{6}
$$

Then, for any $z$,

$$
\tilde{p}_i^t(z) = \sum_{x \in \Psi(z)} \bar{p}_i^t(x)
\tag{7}
$$

Calculating $E[Z_i^t]$ is straightforward:

$$
E[Z_i^t] = \sum_{z \in \Omega} z \tilde{p}_i^t(z)
\tag{8}
$$

To calculate $P_i^t$, the probability that alternative $i$ has the greatest utility, we exploit the discrete distributions and condition on $Z_i^t$ as follows:

$$\begin{aligned} P_i^t &= P\{Z_i^t \succ Z_h^t \ \forall h \neq i \mid S_t\} \\ &= \sum_{z \in \Omega} \tilde{p}_i^t(z) P\{Z_i^t \succ Z_h^t \ \forall h \neq i \mid S_t, Z_i^t = z\} \\ &= \sum_{z \in \Omega} \tilde{p}_i^t(z) \prod_{h<i} P\{Z_h^t < z\} \prod_{h>i} P\{Z_h^t \leq z\} \end{aligned} \quad (9)$$

The cumulative distribution function of $Z_h^t$ is calculated from its probability distribution:

$$P\{Z_h^t \leq z\} = \sum_{s \leq z} \tilde{p}_h^t(s) \quad (10)$$

## 4. Sample Allocation Procedures

In this study we evaluated two sequential lookahead sample allocation procedures and two hybrid procedures that included both uniform allocation and one of the sequential lookahead allocation procedures. Each hybrid procedure includes two special cases: at one extreme, every sample is allocated using the sequential procedure; at the other extreme, every sample is allocated using the uniform allocation procedure. Because the computational effort and state space of dynamic programming would be expensive, it is rational to consider heuristics such as these. The hybrid procedure attempts to get the best features of uniform allocation (less computational effort) and sequential allocation (better sample allocations).

### 4.1. Uniform Allocation

The uniform sample allocation procedure allocates samples to the alternatives and attributes as equally as possible. It can be used as a batch procedure because it can determine a sample allocation before any samples are collected. Let *H* be the total number of samples to be allocated among the *km* alternatives and attributes. Here we consider only the case in which *H* is a multiple of *km*, but the procedure can be extended to other values by specifying a rule for allocating the remaining samples.

**Uniform Allocation**

1. Set $N_{ij} = \dfrac{H}{km}$ for $i = 1, \ldots, m$, $j = 1, \ldots, k$.

**4.2. Sequential Allocation**

The sequential allocation procedures use a one-step lookahead approach to determine the best alternative and attribute pair to sample next. Each procedure consists of *T*+1 stages: stage 0 is the initial state; each subsequent stage begins with obtaining a sample from one alternative and attribute, which leads to a new state. That is, $S_0$ is the state of the decision-maker's beliefs in stage 0, and $S_t$ is the state of the decision-maker's beliefs after obtaining the sample $w_t$ in stage *t*, for *t* = 1, …, *T*.

In stages 0 to *T*-1, the decision-maker uses these beliefs to determine which alternative and attribute to sample at the beginning of the next stage; that is, the decision-maker allocates the next sample. In stage *T*, the decision-maker selects the most preferred alternative. The horizon of this lookahead model is only one stage (one sample).

The two versions of the sequential allocation procedure correspond to the two quality measures (loss functions). In the Sequential I version, the sample allocation is determined by evaluating $E\left[\max_{i=1,\ldots,m} E\left[Z_i^t \mid w_t\right]\right]$ for each possible alternative and attribute and selecting the alternative and attribute that maximizes that quantity (which also maximizes the utility increment, a measure of the value of the information, as discussed in Section 3.1). In the Sequential II version, the sample allocation decision is determined by evaluating the expected values of all $P_i^{t+1}$ if another sample from a specific alternative and attribute pair is obtained and then identifying the alternative and attribute pair that gives the greatest maximum expected value. This approach is equivalent to computing the knowledge gradient (Frazier *et al.*, 2008), and selecting the alternative and attribute pair that gives the greatest knowledge gradient. It is essentially solving a decision tree that has *km* possible choices (one for each alternative and attribute pair), each choice has numerous outcomes (one for each possible magnitude of the sample), and the payoff is the greatest $P_i^{t+1}$. This version of the procedure is a generalization of the sequential sample allocation approach introduced by Leber and Herrmann (2016).

In this section, let $p_{ij}^{t+1}(x_j | w)$ be the posterior probability that the magnitude of attribute *j* of alternative $a_i$ equals $x_j$ if the next sample obtained equals *w*. Let $f_{ij}^{t+1}(w)$ be the maximum $P_i^{t+1}$ if the next sample of attribute *j* of alternative $a_i$ equals *w*.

**Sequential I Allocation**

1. $t = 0$. From the given prior distributions $p_{ij}^0(x_j)$, create the prior distribution $\tilde{p}_i^0(z)$ for the utility of alternative *i*, $i = 1,\ldots,m$.

2. For $i = 1,\ldots,m$, $j = 1,\ldots,k$, do steps 2a and 2b.

2a. For all $w \in X_j$, do steps 2a1, 2a2, and 2a3.

2a1. Calculate the probability $p_{ij}^S(w) = \sum_{y \in X_j} p_{ij}^t(y) p_j^e(w-y)$ that the next sample for alternative $i$, attribute $j$ will be $w$.

2a2. Create the posterior distribution for the true magnitude of alternative $i$, attribute $j$ if the next sample were $w$ (over all $x_j \in X_j$):

$$p_{ij}^{t+1}(x_j \mid w) = \frac{p_{ij}^t(x_j) p_j^e(w - x_j)}{p_{ij}^S(w)}. \tag{11}$$

2a3. Create the posterior distribution $\tilde{p}_i^{t+1}(z)$ for the utility of alternative $i$. For $h \neq i$, $\tilde{p}_h^{t+1}(z) = \tilde{p}_h^t(z)$. Calculate $f_{ij}^{t+1}(w)$ as follows:

$$f_{ij}^{t+1}(w) = \max\{E[Z_h^{t+1}] : h = 1, \ldots, m\} \tag{12}$$

2b. Calculate the expected value $F_{ij}^{t+1}$:

$$F_{ij}^{t+1} = \sum_{w \in X_j} p_{ij}^S(w) f_{ij}^{t+1}(w) \tag{13}$$

3. Determine the alternative $i^*$ and attribute $j^*$ to sample next:

$$i^*, j^* = \arg\max F_{ij}^{t+1} \tag{14}$$

4. Increase $t$ by 1. (Begin next stage.)

5. Draw a sample magnitude $w$ for alternative $i^*$ and attribute $j^*$. Update the distributions for $p_{i^*j^*}^t(x_{j^*})$ and $\tilde{p}_{i^*}^t(z)$. For all other $i$ and $j$, $p_{ij}^t(x_j) = p_{ij}^{t-1}(x_j)$ and $\tilde{p}_i^t(z) = \tilde{p}_i^{t-1}(z)$. If $t < T$, go to Step 2.

6. Select alternative $b = \arg\max_{i=1,\ldots,m} \{E[Z_i^T]\}$.

**Sequential II Allocation**

This procedure follows the same steps as the Sequential I allocation procedure, with the following changes to Steps 2a3, 5, and 6:

2a3. Evaluate $P_h^{t+1}$ for $h = 1, \ldots, m$ using the posterior distributions. Calculate $f_{ij}^{t+1}(w)$ as follows:

$$f_{ij}^{t+1}(w) = \max\{P_h^{t+1} : h = 1, \ldots, m\} \tag{15}$$

5. Draw a sample $w$ for alternative $i^*$ and attribute $j^*$. Update the distributions for $p_{i^*j^*}^t(x_{j^*})$ and $\tilde{p}_{i^*}^t(z)$. For all other $i$ and $j$, $p_{ij}^t(x_j) = p_{ij}^{t-1}(x_j)$ and $\tilde{p}_i^t(z) = \tilde{p}_i^{t-1}(z)$. Evaluate $P_i^t$ for $i = 1, \ldots, m$. If $t < T$, go to Step 2.

6. Select alternative $b = \arg\max\{P_i^T : i = 1, \ldots, m\}$.

The computational effort of these procedures is dominated by Step 2, which must perform Step 2a for $m\sum_{j=1}^{k}|X_j|$ outcomes, and this step must calculate $|X_j|$ quantities for one posterior distribution and $|\Omega|$ quantities for another.

### 4.3. Hybrid Allocation Procedures

This study considered two hybrid allocation procedures that allocate a fixed number of samples using the uniform allocation procedure and then allocate the remaining samples using one of the sequential allocation procedures. The key parameter is $H$, the number of samples to allocate using the uniform allocation procedure. ($H = 0$ is a fully sequential procedure; $H = T$ is the uniform allocation procedure.) The state of the decision-maker's beliefs about the attribute magnitudes after obtaining the first $H$ samples is $S_H$. The sequential allocation procedure then begins at $t = H$. We tested the performance of these procedures at different levels of $H$.

In particular, the Hybrid I allocation procedure uses the Sequential I allocation procedure to allocate the remaining $T$-$H$ samples. The Hybrid II allocation procedure uses the Sequential II allocation procedure to allocate the remaining $T$-$H$ samples.

## 5. Experimental Design

This section describes the simulation study that we conducted to compare the sequential and hybrid sample allocation procedures with different levels of $H$. The procedures were implemented in MATLAB R2016b, and the computations were conducted on a personal computer running an Intel Core i7-6700HQ CPU at 2.60 GHz with 8 GB of RAM.

We generated two problem sets; each problem set included twenty randomly-generated instances. Each instance specified the true magnitudes of every attribute for every alternative. For every attribute $j = 1,\ldots,k$, we used a linear value function that maps the attribute's magnitude to a value in the interval [0, 1]:

$$v_j(x_j) = x_j / \max X_j \tag{16}$$

Let $B_k = 1 + \cdots + k$ for any positive integer $k$. We used two different multi-attribute value functions:

$$v_A(x_1,\ldots,x_k) = \sum_{j=1}^{k} \frac{j}{B_k} v_j(x_j) = \sum_{j=1}^{k} \frac{j}{B_k} \frac{x_j}{\max X_j} \tag{17}$$

$$v_B(x_1,\ldots,x_k) = \sqrt{\frac{1}{k}\sum_{j=1}^{k} v_j(x_j)^2} = \sqrt{\frac{1}{k}\sum_{j=1}^{k}\left(\frac{x_j}{\max X_j}\right)^2} \tag{18}$$

Value function A is an additive value function. Value function B models a preference for compensating solutions (Scott and Antonsson, 2005). Both value functions yield a value in the range [0, 1]. We used two different utility functions:

$$U_0(v) = v \tag{19}$$

$$U_\gamma(v) = \frac{1-e^{-\gamma v}}{1-e^{-\gamma}} \tag{20}$$

Utility function $U_0$ models the risk-neutral case. Utility function $U_\gamma$ models risk aversion, where each instance had a randomly generated $\gamma$ from the range [1, 10]. Note that the problem formulation in Section 3 and the allocation procedures presented in Section 4 do not require any specific forms of the value functions and the utility function; the functions used here were chosen merely for the computational experiment.

All of the prior probability distributions were uniform: $p_{ij}^0(x_j) = 1/|X_j|$.

We compared a total of twelve sample allocation procedures: six levels of $H$ and two decision rules corresponding to the two loss functions. The Sequential I and II procedures are equivalent to the Hybrid I and II procedures with $H = 0$; we will call these the "sequential procedures." The "hybrid procedures" are the Hybrid I and II procedures with $H = 36, 72, 108$, and $144$. The "uniform procedures" are the Hybrid I and II procedures with $H = 180$. In the Sequential I and Hybrid I procedures, the selected alternative at any stage is the one with the greatest expected utility. In the Sequential II and Hybrid II procedures, the selected alternative at any stage is the one with the greatest $P_i^t$. The Hybrid I and Hybrid II procedures make the same sample allocation (the uniform allocation) for $t \leq H$, but they may select different alternatives due to the different decision rules.

We used each sample allocation procedure on each instance; each procedure was run for ten replications. Thus, in each problem set, each procedure was tried 200 times. The levels of $H$ came from the set $\{0, 36, 72, 108, 144, 180\}$.

For both problem sets, for every attribute $j$, $X_j = \{1,\ldots,15\}$, so $|X_j| = 15$. For problem set A, $m = 12$, and $k = 3$. For problem set B, $m = 9$, and $k = 4$. Unless specified otherwise, all probability distributions used in generating instances were uniform over the range of possibilities. We used the following procedure to generate the true attribute magnitudes for each instance.

Step 1. Randomly select an attribute $h$ in the set $\{1, \ldots, k\}$. Randomly select $\alpha$ from the interval $[1, 3]$. For every attribute $j \neq h$, randomly select $c_j \in [0,1]$. Set the "weight" $d_j$ for attribute $j \neq h$ as follows:

$$d_j = \frac{c_j}{\sum_{g \neq h} c_g} \tag{21}$$

Step 2. For each alternative $i = 1, \ldots, m$, repeat Steps 2a and 2b.

Step 2a. For every attribute $j \neq h$, randomly select $x_{ij} \in [0,1)$. For attribute $h$, set $x_{ih}$ as follows:

$$x_{ih} = 1 - \sum_{j \neq h} d_j x_{ij}^{\alpha} \qquad (22)$$

Step 2b. For every attribute $j = 1, \ldots, k$, determine its true magnitude as follows:

$$\mu_{ij} = 1 + \lfloor x_{ij} \max X_j \rfloor \qquad (23)$$

We also chose the error distributions for each attribute so that these were symmetric around 0, and $p_j^e(0)$ was the largest probability in the distribution. For all three problem sets, $E_j = \{-3, -2, -1, 0, 1, 2, 3\}$ for every attribute $j$. Tables 1 and 2 list the error distributions and their standard deviations. When an instance was generated, the error distributions for its problem set were randomly assigned to the attributes (thus, the error distribution of a particular attribute varied by instance).

Table 1. Problem Set A error distributions and standard deviations.

| $p_j^e(e)$ for $e \in E_j$ | Std. Dev. |
|---|---|
| (0.020, 0.116, 0.211, 0.307, 0.211, 0.116, 0.020) | 1.31 |
| (0.080, 0.129, 0.178, 0.227, 0.178, 0.129, 0.080) | 1.68 |
| (0.140, 0.142, 0.144, 0.147, 0.144, 0.142, 0.140) | 1.99 |

Table 2. Problem Set B error distributions and standard deviations.

| $p_j^e(e)$ for $e \in E_j$ | Std. Dev. |
|---|---|
| (0.020, 0.116, 0.211, 0.307, 0.211, 0.116, 0.020) | 1.31 |
| (0.060, 0.124, 0.189, 0.253, 0.189, 0.124, 0.060) | 1.57 |
| (0.100, 0.133, 0.167, 0.200, 0.167, 0.133, 0.100) | 1.79 |
| (0.140, 0.142, 0.144, 0.147, 0.144, 0.142, 0.140) | 1.99 |

## 6. Results

This section first describes the results of the sample allocation procedures in terms of the average opportunity cost of the selected alternatives and the number of correct selections. Section 6.2 discusses why the sequential procedures did not yield superior alternatives. Section 6.3 discusses the computational effort of the procedures.

### 6.1. Opportunity Cost and Correct Selections

For the first decision quality measure, the key performance measure is the average opportunity cost, which equals the difference between the true utility of the selected alternative and the best true utility. If alternative $a_b$ is selected, the opportunity cost is $\xi^* - \xi_b$. For the second decision quality measure, the key performance measure is the number of times that a procedure led the decision-maker to select a truly best alternative (one in $A^*$). For each problem set, instance, replication, and stage, we determined the true utility of the alternative that the decision-maker would select at that point, the opportunity cost of that alternative, and whether that alternative is one of the truly best.

For each sample allocation procedure (level of $H$ and decision rule), problem set, value function, utility function, and stage, we averaged (over the instances and replications) the opportunity cost of the selected alternative and totaled (over the instances and replications) the number of times (out of 200) that a truly best alternative was selected (these are called correct selections).

As shown in Figures 1 to 8, as expected, at each stage the quality of the selected alternatives increases; that is, the average opportunity cost generally decreases and the number of correct selections increases. Recall that the hybrid procedures ($H = 36, 72, 108$, and $144$) and the uniform procedures ($H = 180$) choose the same samples and select the same alternatives in the early stages, which is why they are not distinguishable in these figures. When the hybrid procedures begin to use sequential sample allocation, the quality of the selected alternatives increases greatly at first but then stops increasing. In most cases, at the last stage, the hybrid procedures select equally high-quality alternatives. In some but not all cases, the uniform procedures also yield high-quality alternatives. Likewise, in some but not all cases, the sequential procedures ($H = 0$) also yield high-quality alternatives.

We used ANOVA to determine whether differences in the average opportunity cost were significant at the 5% level. Tables 3 and 4 show the results of comparing the sequential ($H = 0$) and hybrid procedures with the corresponding uniform procedure ($H = 180$). We determined whether differences in the number of correct selections were significant by calculating 95% confidence intervals on the differences following Navidi (2006). Tables 5 and 6 show the results of comparing the sequential ($H = 0$) and hybrid policies with the uniform policy ($H = 180$).

For Problem Set A, as shown in Table 3, the average opportunity cost of the alternatives selected by the hybrid procedures was significantly less than the average opportunity cost of the alternatives selected by the corresponding uniform procedure in every case when $H = 72$ and $H = 108$ and in all but one of the cases when $H = 144$. As shown in Table 5, the number of correct selections by hybrid procedures was significantly more than the number of correct selections by the corresponding uniform procedure in every case when $H = 36, 72, 108$, and $144$. (The number is tallied over 200 total runs.) The quality of the solutions selected by the Sequential II procedure (the Hybrid II procedure with $H = 0$) was significantly better than the quality of the solutions selected by the uniform procedure in three out of four cases. The quality of the solutions selected by the Sequential I procedure (the Hybrid I procedure with $H = 0$) was not better than the quality of the solutions selected by the uniform procedure in any case.

For Problem Set B, as shown in Table 4, the average opportunity cost of the alternatives selected by hybrid procedures ($H = 36$, 72, 108, and 144) was significantly less than the average opportunity cost of the alternatives selected by the corresponding uniform procedure in 12 cases (out of 16) with the risk-neutral utility function. With the risk-averse utility function, the average opportunity cost of the alternatives selected by hybrid procedures was significantly less than the average opportunity cost of the alternatives selected by the corresponding uniform procedure in only three cases (out of 16). Generally, the completely sequential procedures selected significantly worse alternatives, but this depended upon the value function; with value function A, the completely sequential procedures selected alternatives with lower average opportunity cost than they did with value function B. As shown in Table 6, the number of correct selections by the hybrid procedures generally increased as $H$ increased. With the Hybrid II procedure, for $H = 72$, the number of correct selections was significantly more than the number of correct selections by the corresponding uniform procedure in two cases (out of four). The sequential II procedure (Hybrid II with $H = 0$) selected a correct alternative significantly less often than the uniform policy in two cases, both with value function B.

These results confirm that the performance of a procedure generally increases as more samples are obtained. The relative performance of these procedures varies, however. Consider two procedures that use the same decision rule (loss function) but have different levels of $H$, say $H_1$ and $H_2$, with $H_1 < H_2$. When the number of samples is less than $H_1$, then both procedures are selecting samples uniformly, and their performance is the same. When the number of samples is greater than $H_1$ but less than $H_2$, then the first procedure is using sequential sample allocation, and its performance will become better than that of the second procedure. When the number of samples is greater than $H_2$, then both procedures are using sequential sample allocation, and the performance of the second procedure will begin to approach that of the first procedure (in some cases, when $H$ is very large, it may be unable to reach that level because the number of remaining samples is insufficient). Because selecting samples uniformly requires less computational effort than sequential sampling, this suggests that, for a given budget $T$, it may be reasonable to set $H$ to a large number that is less than $T$. This procedure would select most of the samples uniformly and then allocate samples sequentially for the remaining stages to improve the likelihood of correct selection. For example, in Problem Set A, if the total budget were 180 samples, then letting $H = 108$ leads to selecting alternatives significantly better than those selected by the uniform procedure ($H = 180$), and no other procedure performs consistently better.

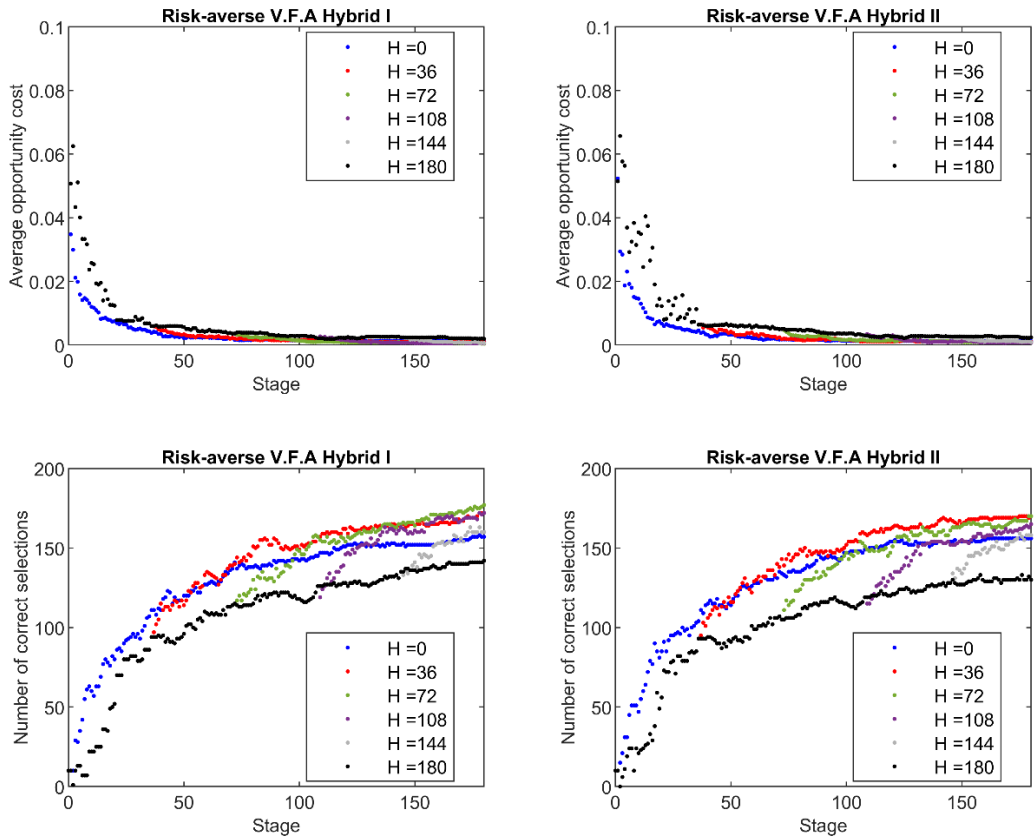

Figure 1. Average opportunity cost and number of correct selections across all replications and instances for Problem Set A with the risk-averse utility function and value function A.

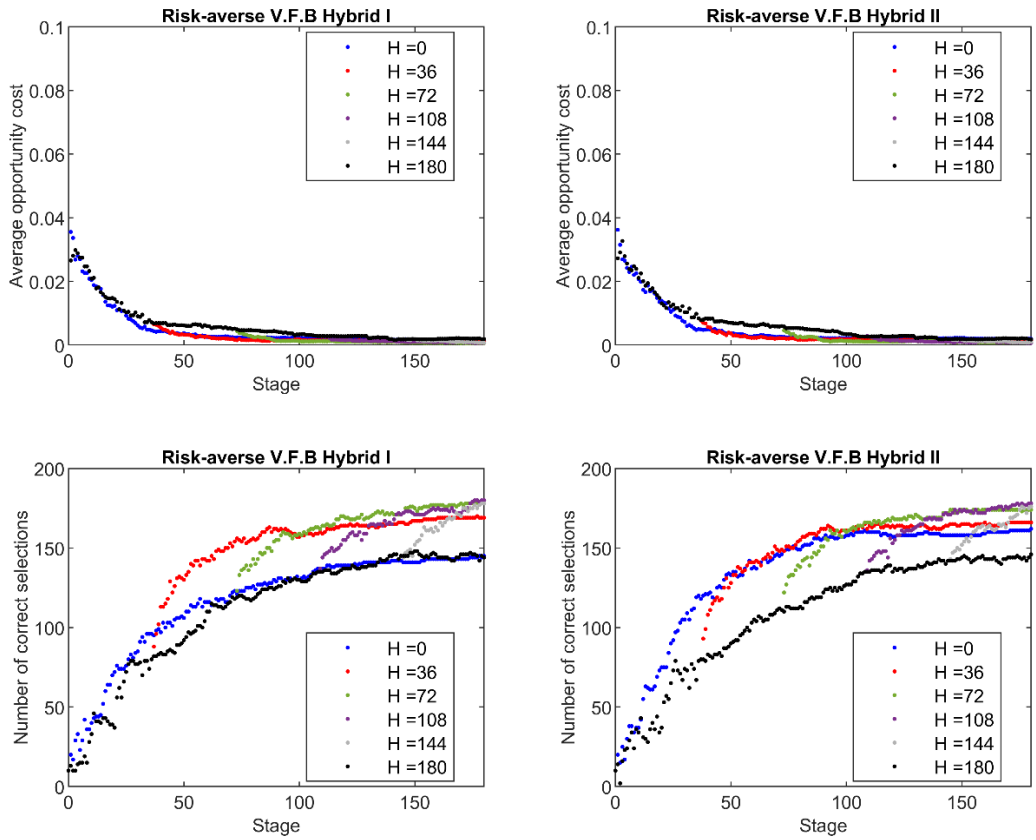

Figure 2. Average opportunity cost and number of correct selections across all replications and instances for Problem Set A with the risk-averse utility function and value function B.

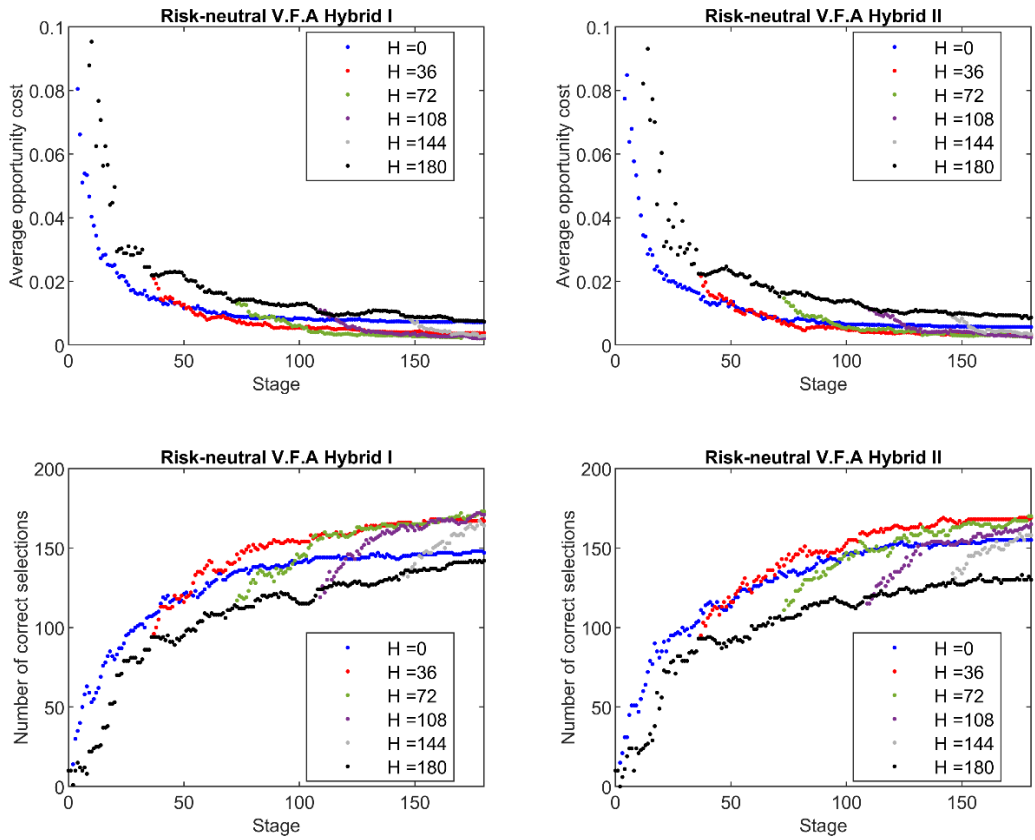

Figure 3. Average opportunity cost and number of correct selections across all replications and instances for Problem Set A with the risk-neutral utility function and value function A.

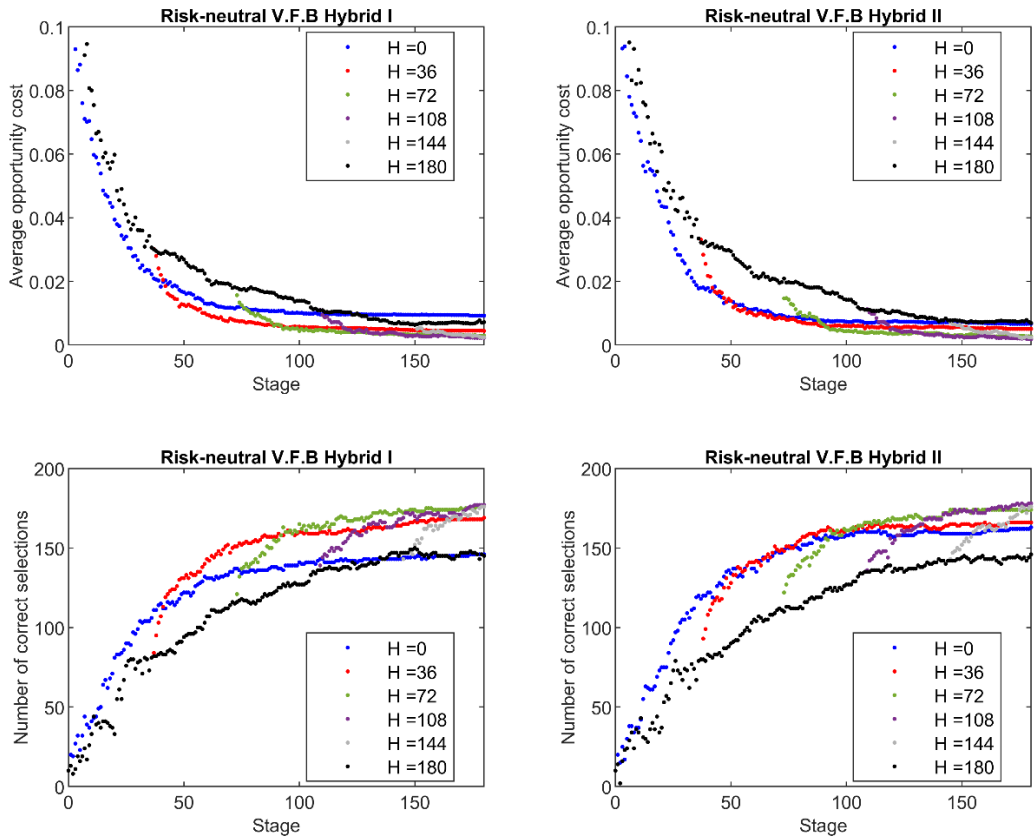

Figure 4. Average opportunity cost and number of correct selections across all replications and instances for Problem Set A with the risk-neutral utility function and value function B.

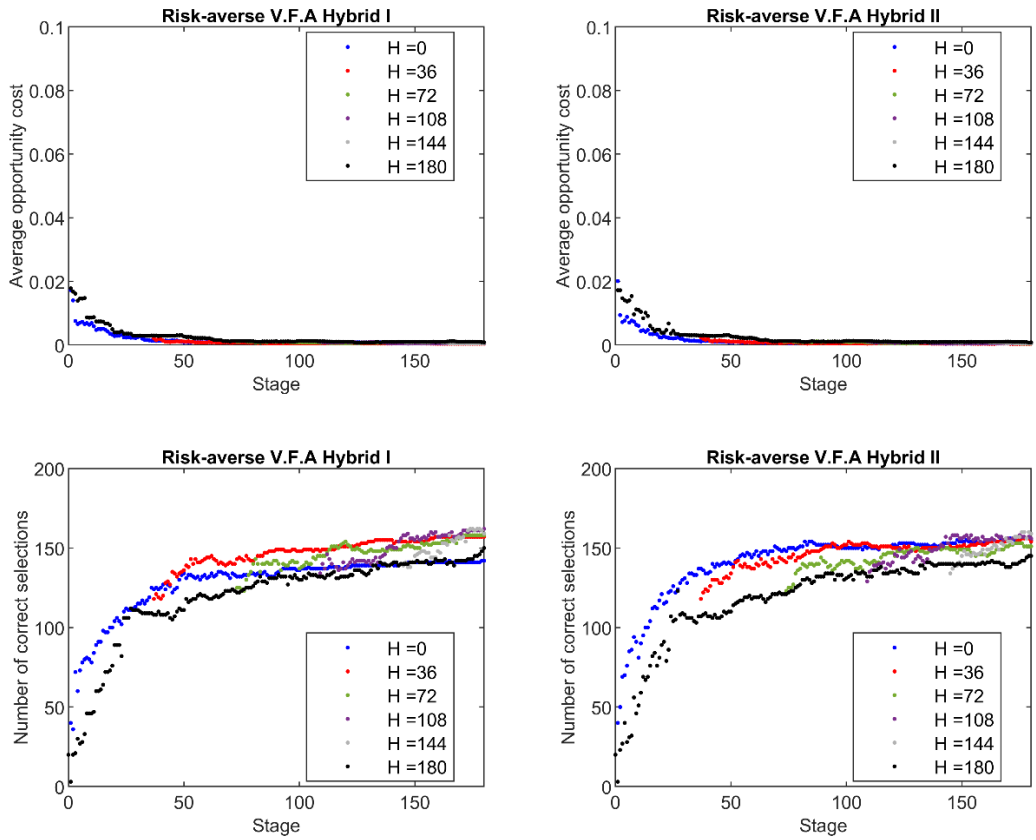

Figure 5. Average opportunity cost and number of correct selections across all replications and instances for Problem Set B with the risk-averse utility function and value function A.

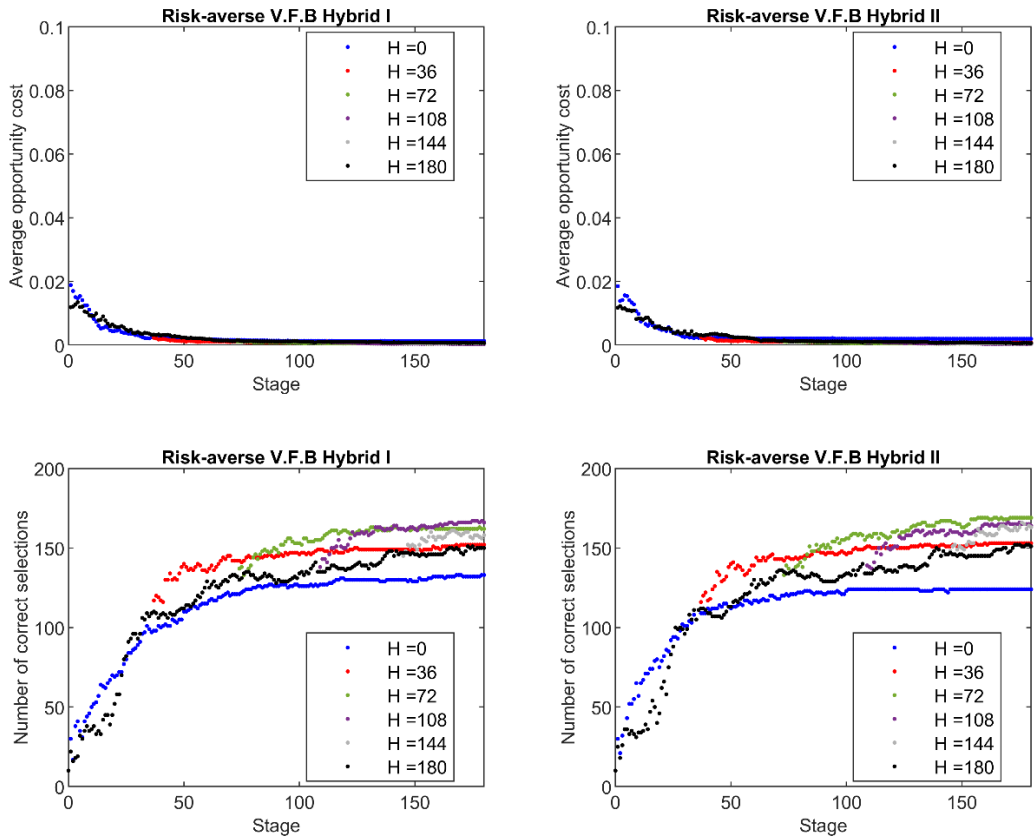

Figure 6. Average opportunity cost and number of correct selections across all replications and instances for Problem Set B with the risk-averse utility function and value function B.

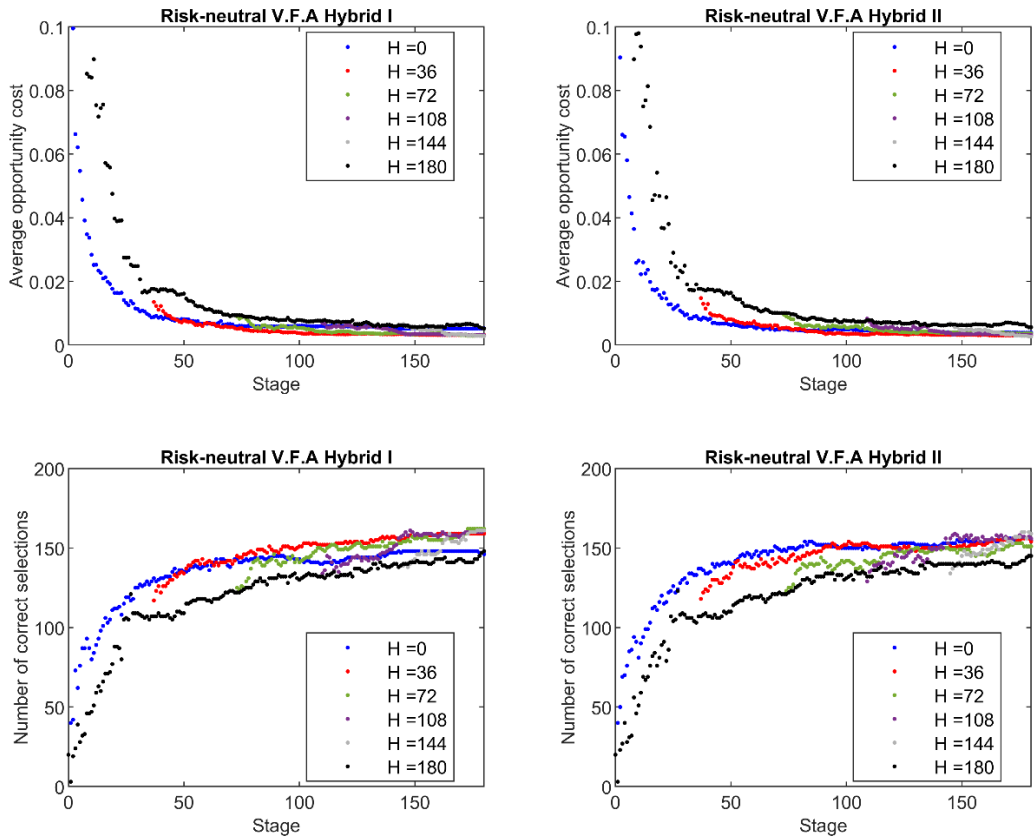

Figure 7. Average opportunity cost and number of correct selections across all replications and instances for Problem Set B with the risk-neutral utility function and value function A.

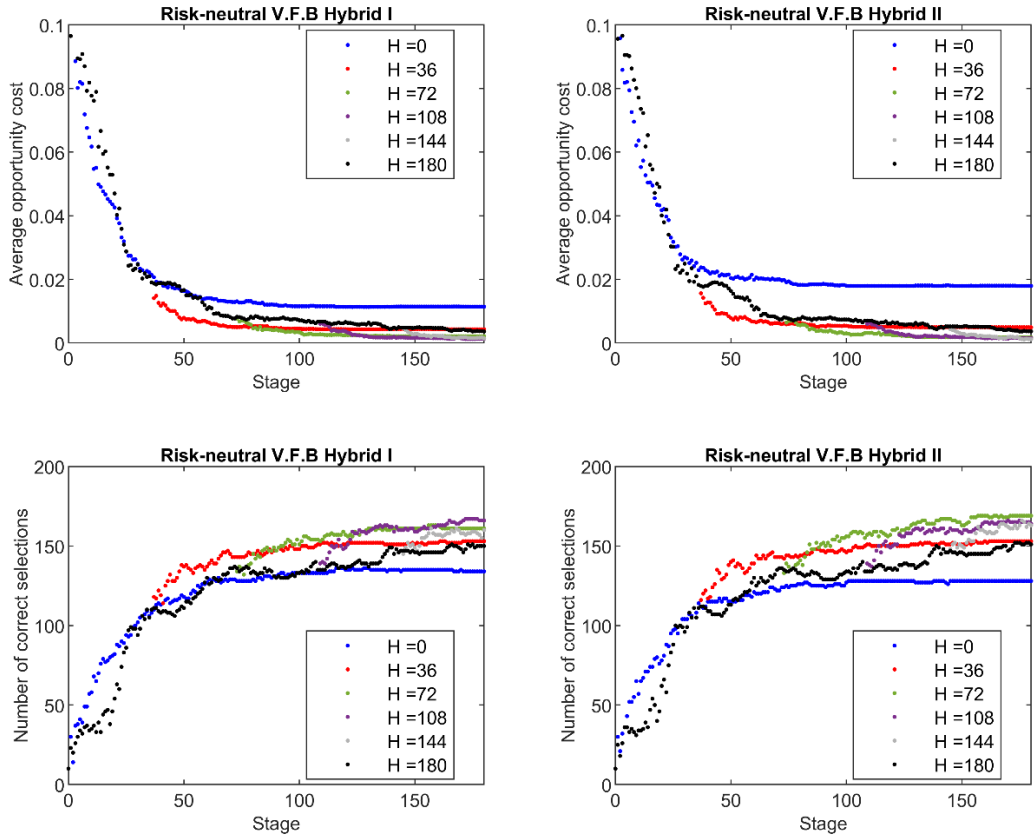

Figure 8. Average opportunity cost and number of correct selections across all replications and instances for Problem Set B with the risk-neutral utility function and value function B.

Table 3. Average opportunity cost of alternatives selected at stage 180 for Problem Set A. Numbers marked with "*" are significantly worse than the average for the uniform procedure ($H = 180$); numbers marked with "†" are significantly better than the average for the uniform policy.

| $H$ | Risk-neutral, Value function A | | Risk-neutral, Value function B | | Risk-averse, Value function A | | Risk-averse, Value function B | |
|---|---|---|---|---|---|---|---|---|
| | Hybrid I | Hybrid II | Hybrid I | Hybrid II | Hybrid I | Hybrid II | Hybrid I | Hybrid II |
| 0 | 0.0072 | 0.0057 | 0.0092 | 0.0066 | 0.0011 | 0.0012† | 0.0015 | 0.0019 |
| 36 | 0.0038† | 0.0032† | 0.0044 | 0.0051 | 0.0010 | 0.0010† | 0.0011 | 0.0014 |
| 72 | 0.0021† | 0.0024† | 0.0029† | 0.0027† | 0.0004† | 0.0008† | 0.0006† | 0.0007† |
| 108 | 0.0021† | 0.0024† | 0.0022† | 0.0019† | 0.0005† | 0.0008† | 0.0006† | 0.0007† |
| 144 | 0.0029† | 0.0033† | 0.0024† | 0.0024† | 0.0006† | 0.0009† | 0.0007† | 0.0010 |
| 180 | 0.0073 | 0.0087 | 0.0072 | 0.0069 | 0.0019 | 0.0024 | 0.0018 | 0.0017 |

Table 4. Average opportunity cost of alternatives selected at stage 180 for Problem Set B. Numbers marked with "*" are significantly worse than the average for the uniform procedure ($H = 180$); numbers marked with "†" are significantly better than the average for the uniform policy.

| $H$ | Risk-neutral, Value function A | | Risk-neutral, Value function B | | Risk-averse, Value function A | | Risk-averse, Value function B | |
|---|---|---|---|---|---|---|---|---|
| | Hybrid I | Hybrid II | Hybrid I | Hybrid II | Hybrid I | Hybrid II | Hybrid I | Hybrid II |
| 0 | 0.0052 | 0.0039 | 0.0114* | 0.0180* | 0.0008 | 0.0005 | 0.0013 | 0.0019* |
| 36 | 0.0031† | 0.0031† | 0.0043 | 0.0049 | 0.0004 | 0.0004† | 0.0006 | 0.0008 |
| 72 | 0.0029† | 0.0036 | 0.0021 | 0.0018† | 0.0005 | 0.0004† | 0.0004 | 0.0004 |
| 108 | 0.0031† | 0.0032† | 0.0013† | 0.0015† | 0.0004 | 0.0005 | 0.0003† | 0.0003 |
| 144 | 0.0028† | 0.0026† | 0.0014† | 0.0012† | 0.0005 | 0.0004 | 0.0004 | 0.0004 |
| 180 | 0.0052 | 0.0056 | 0.0036 | 0.0036 | 0.0008 | 0.0008 | 0.0006 | 0.0006 |

Table 5. Number of correct selections at stage 180 for Problem Set A. Numbers marked with "*" are significantly worse than the average for the uniform procedure ($H = 180$); numbers marked with "†" are significantly better than the average for the uniform policy.

| $H$ | Risk-neutral, Value function A | | Risk-neutral, Value function B | | Risk-averse, Value function A | | Risk-averse, Value function B | |
|---|---|---|---|---|---|---|---|---|
| | Hybrid I | Hybrid II | Hybrid I | Hybrid II | Hybrid I | Hybrid II | Hybrid I | Hybrid II |
| 0 | 147 | 155† | 146 | 163† | 157 | 156† | 145 | 162 |
| 36 | 167† | 169† | 169† | 166† | 172† | 170† | 169† | 166† |
| 72 | 173† | 170† | 176† | 174† | 177† | 170† | 178† | 174† |
| 108 | 171† | 165† | 177† | 178† | 172† | 165† | 180† | 178† |
| 144 | 164† | 158† | 176† | 176† | 160† | 158† | 178† | 176† |
| 180 | 142 | 130 | 145 | 146 | 142 | 130 | 144 | 146 |

Table 6. Number of correct selections at stage 180 for Problem Set B. Numbers marked with "*" are significantly worse than the average for the uniform procedure ($H = 180$); numbers marked with "†" are significantly better than the average for the uniform policy.

| $H$ | Risk-neutral, Value function A | | Risk-neutral, Value function B | | Risk-averse, Value function A | | Risk-averse, Value function B | |
|---|---|---|---|---|---|---|---|---|
| | Hybrid I | Hybrid II | Hybrid I | Hybrid II | Hybrid I | Hybrid II | Hybrid I | Hybrid II |
| 0 | 146 | 155 | 134 | 128* | 142 | 155 | 133 | 124* |
| 36 | 159 | 154 | 153 | 153 | 157 | 154 | 152 | 153 |
| 72 | 162 | 151 | 161 | 169† | 158 | 151 | 162 | 169† |
| 108 | 161 | 156 | 166 | 164 | 162 | 156 | 166 | 164 |
| 144 | 161 | 159 | 160 | 163 | 159 | 159 | 158 | 163 |
| 180 | 148 | 145 | 150 | 151 | 150 | 145 | 150 | 151 |

## 6.2. Sampling Behavior

In many cases, the sequential procedures select higher quality alternatives in the early stages, but the quality stops improving (see, for example, Figures 2 and 6). This occurs because they fail to sample all of the alternative-attribute pairs; instead, they repeatedly sample the same alternative-attribute pair and select the same alternative every stage. Because hybrid procedures use the uniform policy long enough to sample every alternative-attribute pair at least once ($H$ is at least 36), they benefit from this exploration and can exploit it when they begin using sequential sample allocation to choose the next alternative-attribute pair to sample. Eventually, however, these procedures also begin repeatedly sampling the same alternative-attribute pair and selecting the same alternative every stage.

To better understand the behavior of sequential allocation, we examined, at each stage, the alternative-attribute combinations that were sampled and the alternatives that were best and would be selected. We also considered the non-uniformity of the allocation. At stage $t$, after collecting $t$ samples from the 36 combinations, we calculated the quantity $-\sum_{i=1}^{m}\sum_{j=1}^{k}\frac{N_{ij}}{t}\ln\frac{N_{ij}}{t}$, the "entropy" of the set of allocations made in stages 1 to $t$. (For a perfectly uniform set of allocations, the entropy would equal $\ln 36 \approx 3.58$.)

The results show that the sequential procedures ($H = 0$) often did not sample all of the alternative-attribute combinations. Of the 36 combinations, the sequential procedures averaged from 20.7 to 28.8 combinations (the average was taken over all instances and replication). Moreover, it often sampled the same combination repeatedly.

Moreover, although the hybrid allocation procedures ($H = 36, 72, 108, 144$) started by sampling every combination once, the distribution of samples allocated to each combination did not remain uniform. The entropy reached its maximum in stage $t = 36$, remained there through $t = H$, but decreased in subsequent stages. For these hybrid procedures, deviating from the uniform allocation was helpful; after getting at least one sample from every combination, these policies obtained more valuable samples after stage $H$ that lowered the opportunity cost and made selecting the truly optimal alternative more likely. Like the sequential procedures, these procedures were likely, after stage $H$, to sample the same combination repeatedly, which decreased the entropy of the set of sampled combinations as the allocations continued.

The best alternatives were sampled more often after stage $H$. For instance, when $H = 36$, every alternative and attribute received one sample before the sequential procedure was employed; after all 180 stages, the three best alternatives in Problem Set A instances were allocated 18%, 15%, and 11% of the samples; the three worst alternatives were allocated 7%, 6%, and 6% of the samples; in Problem Set B (with nine alternatives), the best alternative was allocated 18% of the samples, and the second and third best alternatives were each allocated at least 13% of the samples.

In value function A, the attributes have unequal weights, and this appears to affect the sampling in the sequential procedure. For example, in Problem Set A (which had three attributes), in cases with value function A and $H = 36$, after all 180 stages, attribute 3 was allocated 37% of the samples; in cases with value function B, attribute 3 was allocated 26% of the samples.

These results suggest that, for many cases, sampling every alternative-attribute combination at least once contributes to the better performance of the hybrid procedures; the sequential procedures failed to do this, which degraded the quality of the selected alternatives. Deviating from the uniform allocation, however, also enabled the hybrid procedures to perform better more quickly than the uniform procedures.

### 6.3. Computational Effort

Tables 7 and 8 list the average computation time needed for all 180 stages for the sample allocation procedures over the different value functions, utility functions, and levels of $H$ (recall that $H = 0$ is fully sequential). The time required decreased as $H$ increased and the number of samples allocated using the sequential procedure decreased. Note that this time includes the time required to update the decision-maker's beliefs and the time needed to determine which alternative and attribute to sample next. The time required to allocate samples for instances in Problem Set B is larger than that for instances in Problem Set A because the instances in Problem Set B have more attributes.

The computational effort is not proportional to the number of stages ($180 - H$) that use the sequential procedure because the computational effort of using the sequential procedure in one stage decreases as the number of samples increases. For example, for Problem Set A, when $T = 180$ and $H = 144$, only 36 stages (20% of the total) are using the sequential procedure, but the total computational effort is much less than 20% of the total when $H = 0$ (fully sequential). This occurs because the prior distributions for the attribute magnitudes are uniform, so, in the early stages, there are many attribute magnitudes with a positive probability $p_{ij}^t(x_j) > 0$, and this increases the number of possible vectors (those with a non-zero probability) that correspond to each utility $z$ (cf. Equations 6 and 7). In the later stages, the samples and the limited range of the measurement error imply that some attribute magnitudes have zero probability $p_{ij}^t(x_j) = 0$, so there are fewer magnitudes that need to be considered and fewer vectors that correspond to each utility $z$. Thus, when $H > 0$, the first $H$ stages use uniform sampling, and the sequential procedure is used only after many attributes have already been sampled, which reduces its computational effort.

These results suggest that using a large $H$ is a reasonable policy because a larger $H$ not only yields higher quality alternatives (as discussed in Section 6.1) but also reduces the computational effort required to make the sample allocation decision.

Table 7. Average computation time (seconds) for Problem Set A. The times are the total time for all 180 stages, averaged over the instances and replications.

| $H$ | Utility function A | | Utility function B | | Utility function C | | Utility function D | |
|---|---|---|---|---|---|---|---|---|
| | Hybrid I | Hybrid II | Hybrid I | Hybrid II | Hybrid I | Hybrid II | Hybrid I | Hybrid II |
| 0 | 87.7 | 154.1 | 77.0 | 323.4 | 82.8 | 154.0 | 77.5 | 322.7 |
| 36 | 14.0 | 64.1 | 13.6 | 182.7 | 13.8 | 63.9 | 13.3 | 183.0 |
| 72 | 6.6 | 41.7 | 6.6 | 123.2 | 6.5 | 41.7 | 6.4 | 123.4 |
| 108 | 3.1 | 25.6 | 3.1 | 77.1 | 3.1 | 25.7 | 3.1 | 77.3 |
| 144 | 1.3 | 12.3 | 1.3 | 37.3 | 1.3 | 12.3 | 1.3 | 37.4 |
| 180 | 0.1 | 0.2 | 0.1 | 0.6 | 0.1 | 0.2 | 0.1 | 0.6 |

Table 8. Average computation time (seconds) for Problem Set B. The times are the total time for all 180 stages, averaged over the instances and replications.

| H | Utility function A | | Utility function B | | Utility function C | | Utility function D | |
|---|---|---|---|---|---|---|---|---|
| | Hybrid I | Hybrid II | Hybrid I | Hybrid II | Hybrid I | Hybrid II | Hybrid I | Hybrid II |
| 0 | 828.2 | 1435.8 | 745.5 | 1576.1 | 737.4 | 1405.0 | 836.5 | 1555.4 |
| 36 | 77.6 | 316.7 | 76.6 | 531.3 | 73.6 | 312.5 | 73.3 | 532.7 |
| 72 | 25.0 | 195.4 | 24.7 | 339.7 | 24.7 | 194.6 | 24.7 | 340.7 |
| 108 | 10.2 | 120.3 | 10.5 | 210.4 | 10.2 | 118.9 | 10.3 | 211.0 |
| 144 | 4.1 | 57.7 | 4.0 | 101.1 | 4.0 | 57.0 | 4.0 | 101.3 |
| 180 | 0.7 | 1.4 | 0.7 | 2.0 | 0.7 | 1.4 | 0.7 | 2.0 |

## 7. Conclusions and Future Work

This paper discussed the problem of allocating samples (measurements) to individual attributes in order to obtain information to make a multi-attribute selection decision. In this study, the decision-maker updates his beliefs based on the samples obtained and selects the alternative that maximizes the expected decision quality; we considered two decision quality measures that are equivalent to minimizing opportunity cost and 0-1 loss. The decision-maker seeks to improve the quality of the decision (minimize the expected loss). In this problem, the decision-maker is not considering whether or not to obtain the samples, so the cost of obtaining the samples is not relevant, and the monetary value of the information is not needed. Although it is possible to express the problem in monetary terms, the problem formulation is more general; monetary values are not necessary.

The paper presented novel hybrid sample allocation procedures that utilize two procedures: a uniform allocation approach and a sequential approach. Although specific value functions and utility functions were used in our computational experiments, the approach can be used with any value functions and utility function. For the instances that we considered, the hybrid procedures were more likely to lead the decision-maker to a correct selection and yielded better alternatives (those with lower average opportunity cost). The sequential procedures, although sometimes better than the uniform procedures, did not select better alternatives than the hybrid procedures, and they required the most computational effort. The uniform procedures required the least computational effort, but their performance was inferior to the best hybrid procedures, which benefitted from both the uniform procedure's exploration, which generated information about every alternative and attribute, and the sequential procedures' ability to exploit this information to identify the best alternatives.

These results suggest that, when multiple measurements can be obtained, a two-phase approach is effective: (1) allocate a majority of the samples (measurements) in a uniform way to every alternative-attribute pair, obtain these samples, and update the decision-maker's beliefs about the magnitudes of every attribute; then (2) use the sequential, lookahead procedure to determine the next alternative-attribute pair to sample, obtain that sample, update the decision-maker's beliefs about that alternative-attribute pair, and repeat until no more samples can be obtained.

The learning procedures presented in this paper can be applied to a wide variety of multi-attribute decision situations; they make no assumptions about the distributions of the errors or the decision-maker's beliefs. The sequential sample allocation procedure is a look-ahead procedure that finds an optimal solution to the one-sample (single-stage) problem (Frazier *et al.*, 2008). Moreover, because the probability distributions are discrete, the lookahead procedures do not need to use approximations when calculating the probability that an alternative is truly best.

Thus, this paper adds to the existing set of procedures available for learning problems (cf. Powell and Ryzhov, 2012).

Theoretical analysis of the sequential procedure is ongoing but beyond the scope of this paper. Future research should investigate the benefit of calculating and employing bounds on the expected gain of a sample and consider more problem sets to determine if the performance of these procedures is robust.

**Acknowledgements**